# PERFORMANCE ANALYSIS OF VoIP TRAFFIC OVER INTEGRATING WIRELESS LAN AND WAN USING DIFFERENT CODECS

Ali M. Alsahlany[1]

[1] Department of Communication Engineering, Al-Najaf Technical College, Foundation of Technical Education, Iraq

*ABSTRACT*

*A simulation model is presented to analyze and evaluate the performance of VoIP based integrated wireless LAN/WAN with taking into account various voice encoding schemes. The network model was simulated using OPNET Modeler software. Different parameters that indicate the QoS like MOS, jitter, end to end delay, traffic send and traffic received are calculated and analyzed in Wireless LAN/WAN scenarios. Depending on this evaluation, Selection codecs G.729A consider the best choice for VoIP.*

## 1. INTRODUCTION

During the recent years, there is a growing trend in real time voice communication using Internet protocol (IP). Voice over Internet Protocol (VoIP) is a technology that allows delivery of voice communications over the Internet or other packet switched networks rather than the traditional Public Switched Telephone Network (PSTN). Many VoIP applications are available on the internet: Skype, Viber, Tango, and Yahoo messenger. All of these applications provide good quality, and free calls. In VoIP the analogue voice signal of the transmitter is converted into digital format before compression and encoding it into a stream of IP packets for transmission to the receiver over IP network. At the receiving end, Digital to Analogue Converter (DAC) works on regenerating the original analogue voice signal after reassemble received IP packets in order and processing it [1].

There are many authors have worked on various Quality of Service (QoS) parameters using different service classes in different network types. A study was conducted on various quality parameters impacting on the VOIP service performance. The study refers that these parameters of QoS are required to increase the performance of a VoIP. Researchers in [2], have compared the performance of the VoIP in both Ethernet LAN (802.3) and Wireless LAN (IEEE 802.11). They examine how VoIP performs in two different network setups and analyzes the results obtained using OPNETsimulator. They also examine the optimization of IEEE 802.11e for QoS using the priorities to provide real time service for VoIP. Various QoS parameters like throughput and average delay for VoIP using different protocols are analyzed in Ref. [3] using OPNET simulator. Simulation results show that the Optimized Link State Routing protocol has better performance in terms of throughput and average delay. Similar analysis has been conducted in [4] to analyze the QoS of VoIP deployment over WiMAX network and compare the performance obtained over various service classes. In [5] authors discussed a handoff mechanism effect on VoIP traffic in Wireless LANs for intra and inter mobility. The study was oriented towards the assessment of the quality of the voice traffic, the packet delays, and delay jitter during the handover operation. Atiur et.al. [6], used an analytical model to estimate the VoIP call capacity of IEEE 802.11g wireless LAN to support single hop networks and multi-hop VoIP services. The authors discovered that, the number of hops between VoIP transmitters in IEEE 802.11g wireless LAN has effects on call quality. Hussein et al. [7], conducted a comparison between different queue algorithms. The authors found that Priority queue and Weight Fair Queuing algorithms are the most appropriate to improve QoS for VoIP. In [8] authors used VoIP as the multimedia measurement to identify the main limitations for enhancing QoS in Wireless LANs, compared to wired networks.

However, none of them have presented accurate approximations to predict the performance of QoS for VoIP when various audio codec schemes are used. Also, most of these studies were performed in a small distance network area without taking real WAN as a one of real application environment. In his paper, the OPNET simulator used to implement the proposed VoIP Network. The simulation model for the proposed network is a private network for a company has two locations located at two different countries around the world in order to simulate the communications between two locations as a long distance and the same location as a local. The major factors that affect on the QoS for VoIP such as: mean opinion score (MOS), average end to end delay, jitter, traffic send, and traffic received are analyzed according to International Telecommunication Union (ITU) standards.

A comparison was carried out between different codecs (G.711, G.729A and G.723.1) which are the most appropriate to improve QoS for VoIP.

The rest of this paper is organized as follows. Section 2 focus on VoIP codec schemes. The model design and configuration networks are described in Section 3. Section 4 presents simulation results and analysis. Finally, Section 5 concludes the paper.

## 2. VoIP Codecs

A codec is the term used for the word coder-decoder, converts of analog audio signals into compressed digital form for transmission and then back into an uncompressed audio signal for the reception. There are different codec types based on the selected sampling rate, data rate, and implemented compression. the most common codecs used for VoIP applications are G.711, G722, G723, G726, G728, G729A, etc. each of which varies in the sound quality, the bandwidth required, the computational requirements, encoding algorithm and coding delay [3, 9, 10]:

In our evaluations codecs of ITU standards for audio compression and decompression are used. Following table lists some features of the most common codecs: G.711, G.723.1, and G.729A.

Table 1: Features of the most common codecs: G.711, G.723.1 and G.729A

| IUT-T Codec | Algorithm | Codec Delay (ms) | Bit Rate (Kbps) | Packets Per Second | IP Packet Size (bytes) | Comments |
|---|---|---|---|---|---|---|
| G.711 | pulse-code modulation (PCM) | 0.375 | 64 | 100 | 120 | Delivers precise speech transmission. Very low processor requirements |
| G.723.1 | Multipulse maximum likelihood quantization/algebraic-code excited linear prediction | 97.5 | 5.3 | 33 | 60 | High compression with high quality audio. Can use with dial-up. Lot of processor power |
| G.729A | Conjugate Structure – Algebraic Code Excited Linear Prediction (ACELP) | 35 | 8 | 100 | 50 | Excellent bandwidth utilization. Error tolerant |

## 3. NETWORK MODEL DESCRIPTION

Our simulation approach uses OPNET simulator for network modeling. OPNET simulator is an authoritative communication system simulator developed by OPNET Technologies.

This simulation model was run in three different scenarios to evaluate the difference in performance and to determine the best audio encoding schemes of utilizing VoIP over integrating Wireless LAN/WAN. All the scenarios follows the same structure and the same topology. The basic difference in all the scenarios is with the configuration of the corresponding application and the profiles. As one scenario is implemented with the codec G.711, G.723.1, and G.729.1. Various comparisons conducted to find the values of various parameters.

The simulation model for the network under study is illustrated in Figure 1. The Application Config, Profile Config, Mobility Config and QoS Attribute Config are included in the model. The Wireless LANs has been considered as the components of the whole network and its nodes are using VoIP services.

A frame relay model supporting Permanent Virtual Circuit (PVC) implemented for the WAN. At each end of the frame relay we had subnets. Inside these subnets was a Wireless LAN environment. The Wireless LAN environments consisted of 3 Routers and 5 workstations in each Router. All the Routers connected together using a Switch. At wireless subnet_0 as shown in Figure 2, Two Servers were connected to the Ethernet Router titled VoIP_Server and Server. The VoIP_Server handled the VOIP traffic and the Server handled the FTP and HTTP traffic. All The Wireless LAN Routers were connected to the frame relay via an Ethernet Router that connected the Switch. The WAN and Wireless LAN were both star topologies.

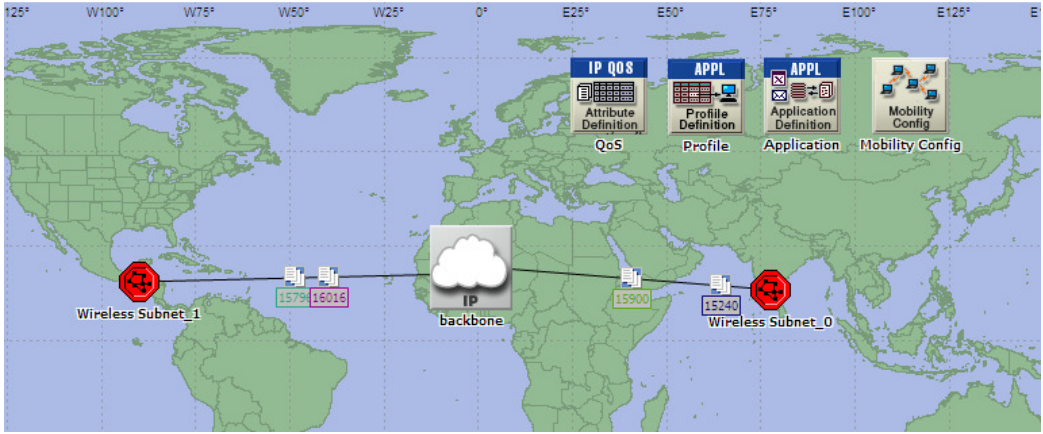

Figure 1: The Simulation Network Model

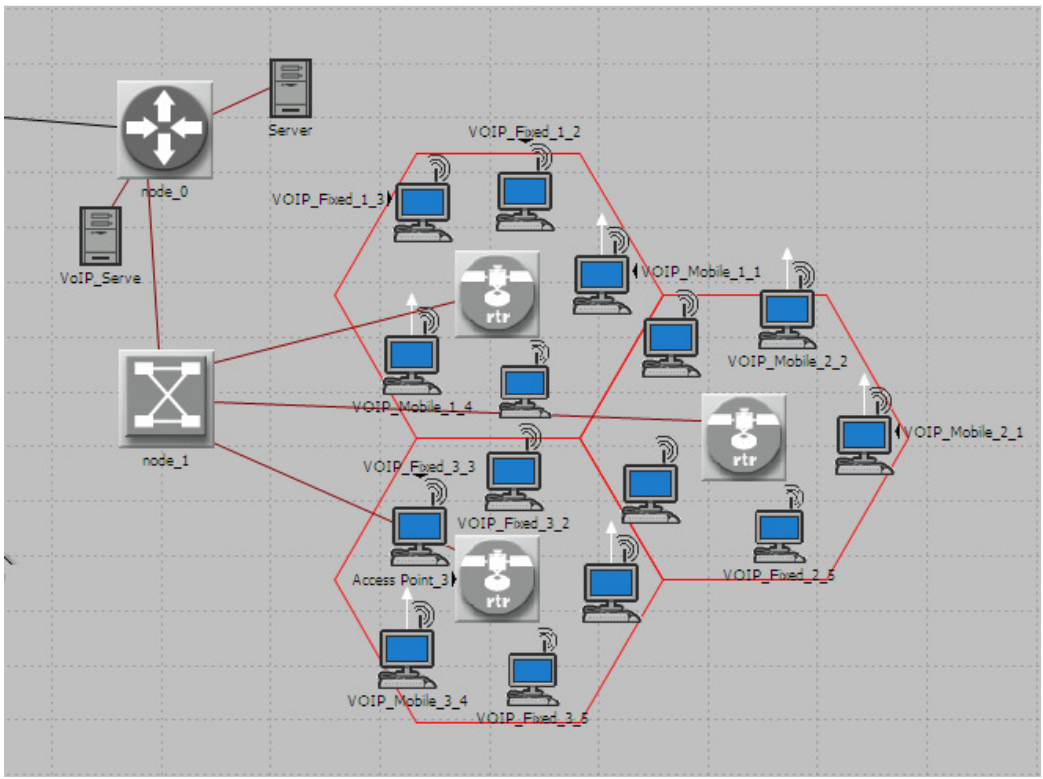

Figure 2: The wireless subnet_0

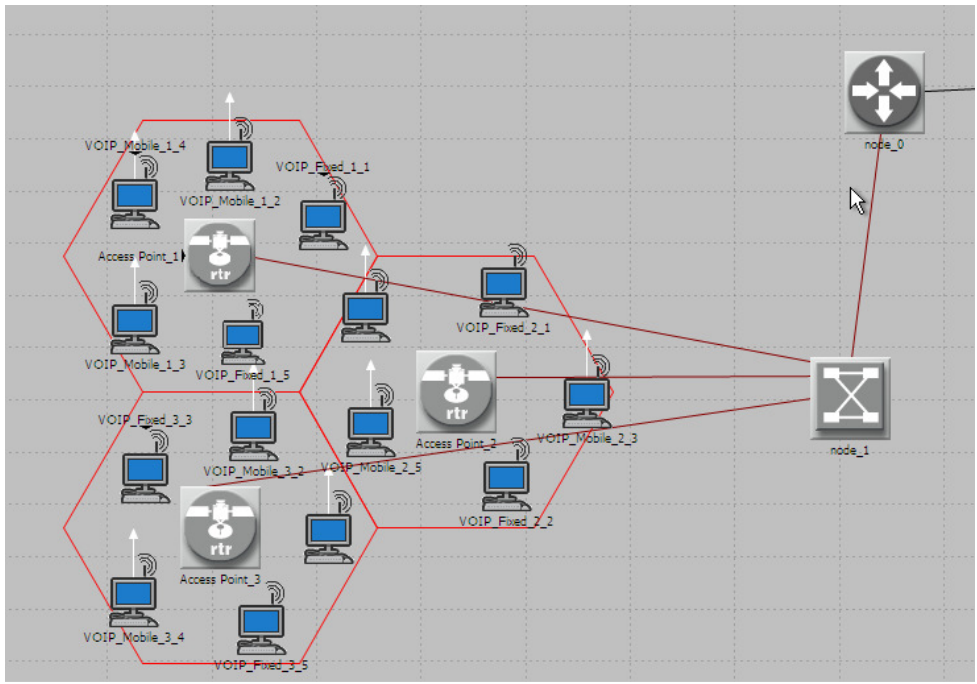

Figure 3: The wireless subnet_1

## 3.1 BASIC PARAMETER

The Point to Point WAN was connected directly to the Ethernet Router in each subnet using a DS0 line. The model name of the link was 'point_to_point_link_adv'. The Wireless LAN environment used 1Gbps Ethernet cable for all the wired connections to the Router, Switch, and Servers. The OPNET model name for the Servers was 'ethernet_server'. Each Wireless LAN environment contains a Router and 2 fixed wireless workstations and 3 moving wireless workstations in each Router. All these wireless nodes use VoIP services. The Routers and workstations were using 802.11g running at a data rate of 54 Mbps. There were a total of 15 workstations in each subnet. The workstations were generating traffic across the WAN and the Wireless LAN environment to simulate a real office environment. The Routers OPNET model name was 'wlan_ethernet_slip4_adv. The fixed VoIP workstation model and the moving VoIP workstation model name were 'wlan_wksn_adv'.

All the parameters using in the Wireless LAN environment have the same value in the same subnet. The Wireless LAN attributes are shown in Table 2.

Table 2: Parameters used in simulation

| Parameter | Value |
|---|---|
| Physical characteristic | IEEE 802.11g |
| Data Rate | 54 Mbps |
| Roaming Capability | Enable |
| Transmit Power | 0.005 W |
| Power Threshold | -95 db |
| Short Retry Limit | 7 |
| Long Retry Limit | 4 |

International Journal of Wireless & Mobile Networks (IJWMN) Vol. 6, No. 3, June 2014## 3.2 APPLICATION, PROFILE, AND MOBILITY CONFIGURATIONS

Application Definitions were where the applications were created for traffic generation on the network. There are 3 applications used in each scenario. These are VoIP, FTP and web browsing applications. The important one was VoIP. In the Voice category IP Telephony selected for configuration VoIP. In the VoIP application, voice attributes setup as codec scheme G.711 for scenario 1; G.723.1 for scenario 2; and G.729A for scenario 3. G.729A operates at a bit rate of 8 Kbps, G.723.1 operates at a bit rate of 5.3/6.3 Kbps and G.711 operates at a bit rate of 64 Kbps. For the QoS scenarios the VoIP table type of service (ToS) was changed to 8 to give voice priority. Other parameters that were set for this project were file transfer and web browsing. For each of these applications mentioned there was a heavy usage parameter set for each one.

Profile Configuration describes the activity applications used by users through a period of time. The FTP_Profile was using VoIP, FTP, Web Browsing and File Transfer with light usage parameters. VoIP_Profile was only using VoIP applications, these were the dedicated IP phones. All the Profiles configured to run simultaneously to allow more than one application to operate at the same time.

The Profile Definition for VoIP traffic is configured that the first VoIP call is starting at 100 second (60 seconds offset time + 40 seconds start time) of the simulation, after that one VoIP call is added after 1 second. The addition of VoIP calls is prepared by repeating the voice application in the Profile Definition. This procedure continues till the end of the simulation.

Mobility Configuration setup the trajectory for the wireless mobile nodes in the Wireless LAN environments. There were two vector mobility profiles. The differences in the profiles were mobility speed.

## 4. SIMULATION RESULTS AND ANALYSIS

The following figures are obtained after collecting statistics by using OPNET Modeler simulation tool. Each figure shows a comparative picture of the three scenarios. All the three scenarios are using a different audio codec scheme such as G.911, G.923 and G.929. After successfully running the simulation, The result shows the impact of different codecs on different QoS parameters in a VoIP network. Following are the figures that show different QoS parameters like MOS, Voice packet end to end delay (sec), Voice jitter (sec), Voice traffic sends (packet/sec) and Voice traffic received (packet/sec).

The most widely used QoS metric in VoIP applications is MOS. The MOS value describes the voice perception quality. The average MOS value for the three codecs is represented in Figure 4. Codecs G. 711 and G. 729A have acceptable MOS values 3.685 and 3.067, respectively. On the other hand, the MOS value for G. 723.1 is 2.557 which indicates that the quality of service is poor if this codec used.

84

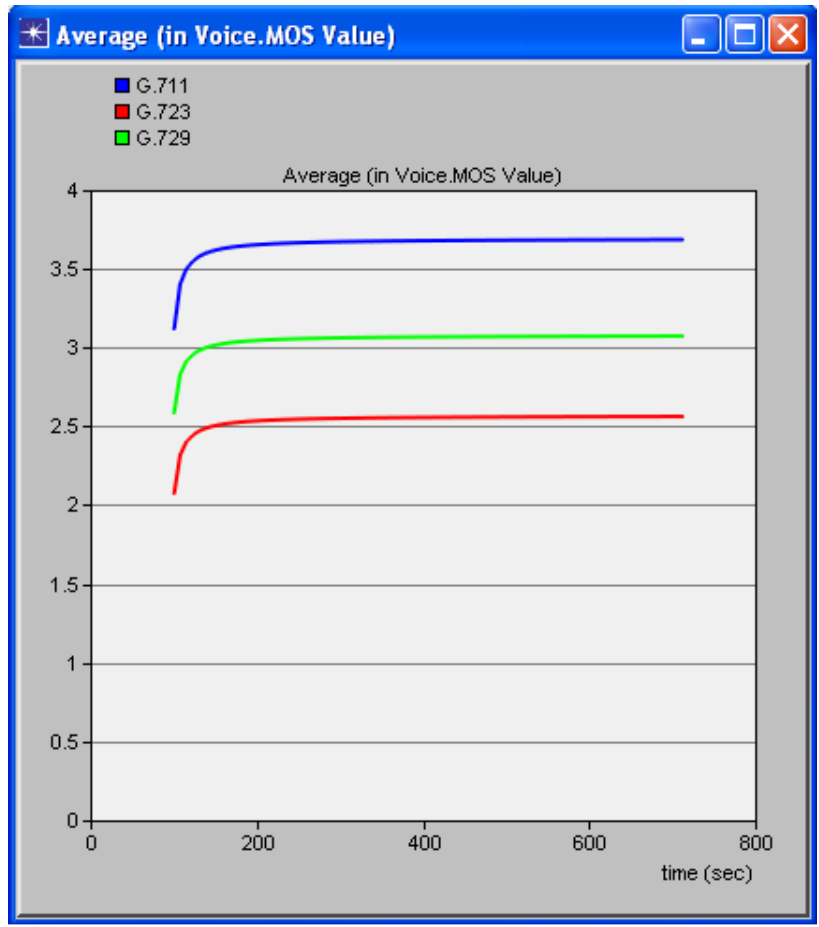

Figure 4: Average voice MOS under various codecs

Average end to end delay metric is shown in Figure 5. G.729A presents the best performance with respect to other codecs. These results are due to transfer rate and packet size. The low packets transfer and the larger packet size, the more time is required to process them. The relatively high transfer rate (8 kbps) and low packet size (20 bytes) for G.729A make G.729A the ideal codecs. Otherwise, G.723.1 and G.711 suffered highest delay than G.729A for the reason that it has the lowest bit rate (5.3 kbps for G.723.1) and larger packet size (160 bytes for G.711). In turn, the end to end delay is increased with transfer rate and packet size.

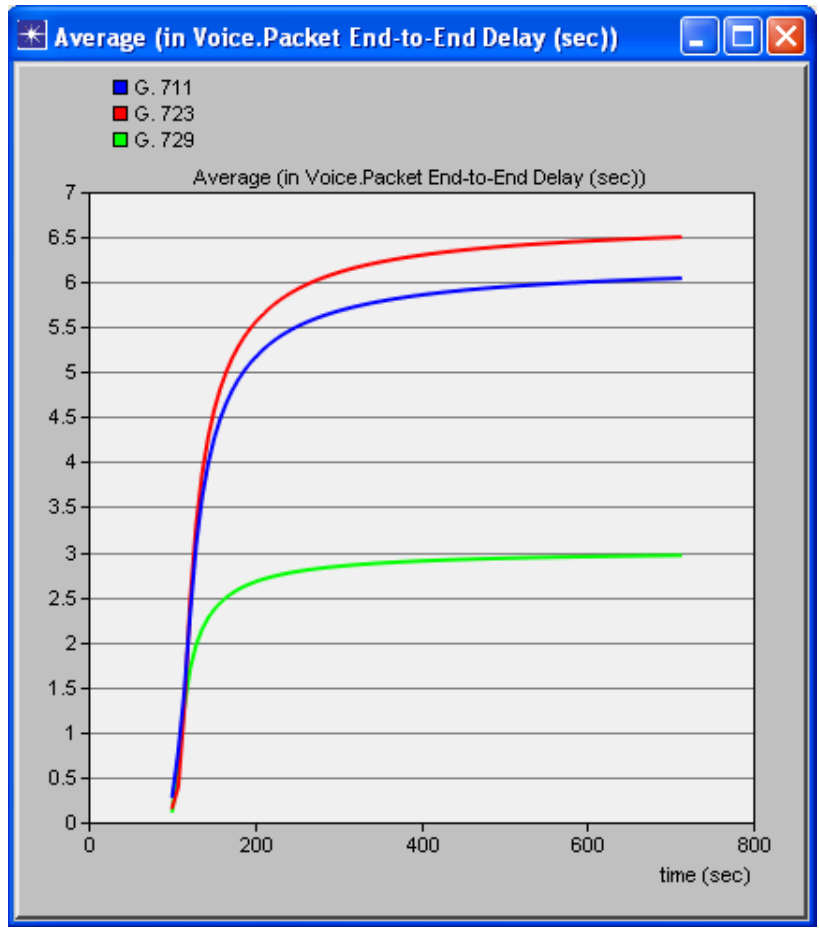

Figure5: Average voice packet End to End delay (sec) under various audio codecs

Figure 6 describes the average voice jitter comparison using different codecs. From the Figure, the variation of the codec G.729A is minimum and approximately constant throughout the simulation. The average voice jitter variation in case of codec G.723.1 is higher than the other two codecs at the earlier time of simulation, but after some time it falls down. The jitte variation in case of G.711 lies between two other audio codecs. The voice jitter threshold for smooth communication in VoIP network is about 1 ms [11] so audio codec G.729A gives better results than audio codecs G.711 and G.723.1 respectively. So there is a high increase in jitter as audio codecs G.711 and G.723.1 are added to the network. This increase in voice jitter makes the voice difficult to understand due to arriving packets at different time . The use audio codec G.729A will make the jitter less and best performance of VoIP application in Integrating Wireless LAN and WAN.

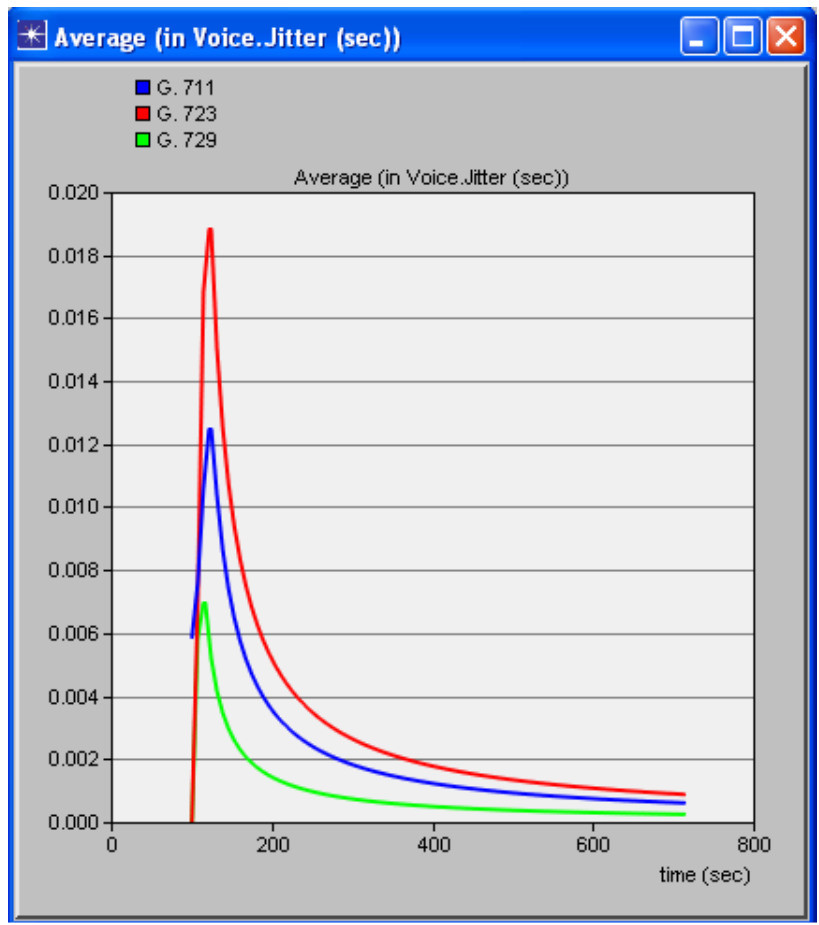

Figure 6: Average Jitter under various audio codecs

Average voice traffic sent and received is presented in Figure 7. Any network to be more efficiency these two traffics must be equal. The traffic received by the network with codec G. 729 is less deviated from the traffic sent comparatively with codec G.723.1 and G.711. This analysis indicates that the noise added in the G.729A network is less when compared to the other networks, so this codec is more efficient.

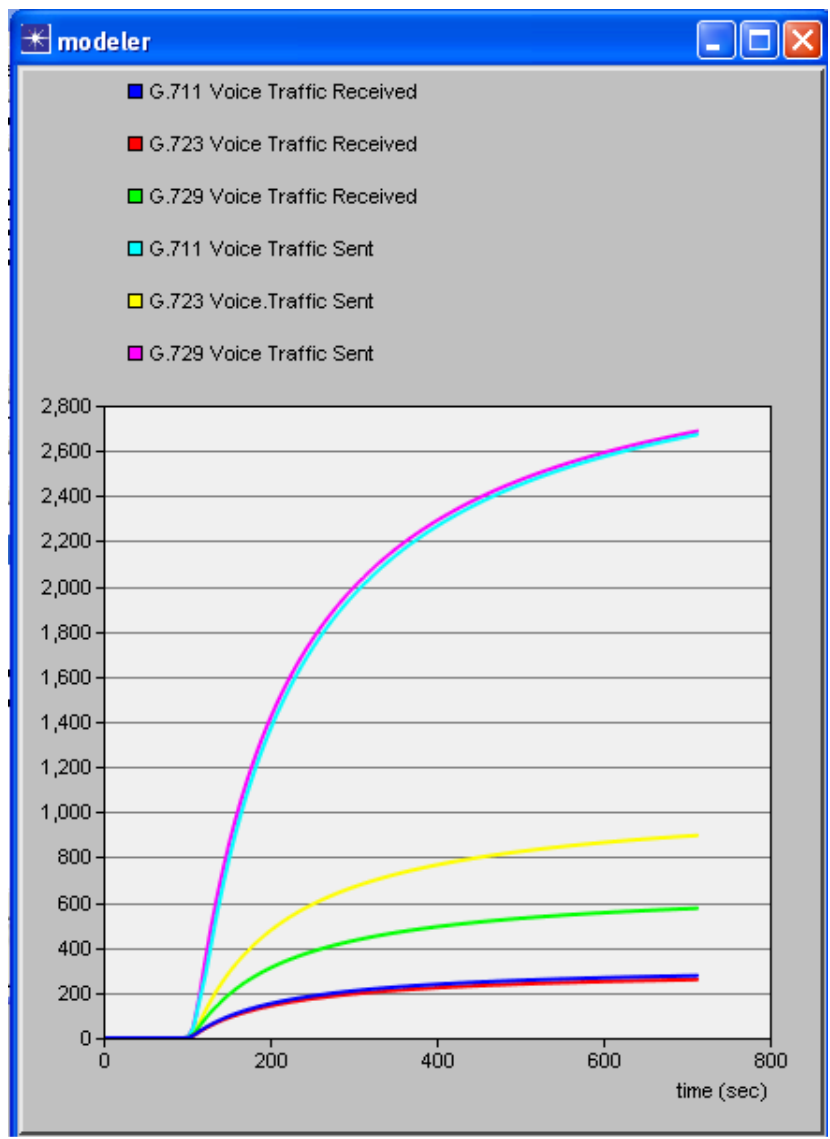

Figure 7: Average voice traffic sent and received under various audio Codecs.

## 5. CONCLUSION

In this paper, analysis and evaluation of the QoS performance for VoIP traffic under various voice codecs was carried out. The use of codecs appropriately is very important in the implementation of VoIP to generate maximum QoS value. The result shows a selection of G.729A codec in a simulation gives a significant result for the performance of VoIP that codec G.729A has acceptable MOS value and less deviation of received to transmit packet as compared to G.711 and G.723.1 also average delay like end to end delay and Voice jitter is lesser in codec G.729A as compared to the other two referenced codecs.

**Author**

Ali M. Alsahlany (alialsahlany@yahoo.com) has got his MSc in communications engineering from Basrah University in 2012. He is now working as a lecturer for computer networks and wireless communications engineering in Communication Engineering Department at Foundation of Technical Education. His area of interest includes Computer Networks, Wireless and Mobile Networks, Optical Communicationand , and Data Communications.

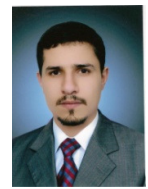